\documentclass[12pt]{iopart}
\expandafter\let\csname equation*\endcsname\relax
\expandafter\let\csname endequation*\endcsname\relax
\usepackage{iopams}
\usepackage{amsmath}
\usepackage{bm}
\usepackage{graphicx}
\usepackage{amssymb}
\usepackage{color,soul}
\usepackage[T1]{fontenc}
\usepackage{ae,aecompl}

\begin{document}

\title[Bifurcation analysis of a TaO memristor model]{Bifurcation analysis of a TaO memristor model}

\author{Y.~V.~Pershin}
\ead{pershin@physics.sc.edu}
\address{Department of Physics and Astronomy, University of South Carolina, Columbia, South Carolina
29208, USA}

\author{V. A. Slipko}
\ead{vslipko@uni.opole.pl}
\address{Institute of Physics, Opole University, Opole 45-052, Poland}

\vspace{10pt}
\begin{indented}
\item[]June 2019
\end{indented}

\begin{abstract}
This paper presents a study of bifurcation in the time-averaged dynamics
of TaO memristors driven by narrow pulses of alternating
polarities. The analysis, based on a physics-inspired model, focuses on the stable fixed points
and on how these are affected by the pulse parameters. Our main finding is the identification of a
driving regime when two stable fixed points exist simultaneously. To the
best of our knowledge, such bistability is identified in a
single memristor for the first time. This result can be readily tested experimentally, and is
expected to be useful in future memristor circuit designs.
\end{abstract}

\maketitle

\newpage

\section{Introduction}
The past decade has witnessed a growing interest in memristive devices and systems~\cite{chua76a} (memristors), both from a fundamental point of view and with a view to applications ~\cite{pershin11a,Yang13a,adamatzky2013memristor,ielmini2015resistive}.
However, even today there are still a lot of open questions to be answered concerning, for instance,
the formulation of predictive models~\cite{biolek2013reliable,Biolek18a},
the fabrication of device structures with desirable and repeatable switching characteristics, their integration with CMOS circuitry~\cite{Xia09a}, and even the foundations of the memristor concept~\cite{di2013physical,vongehr2015missing,sundqvist2017memristor}.  By definition~\cite{chua76a}, a voltage-controlled memristive system is a two-terminal resistive device with memory described by
\begin{eqnarray}
I(t)&=&G\left(\boldsymbol{x}, V_M \right)V_M(t), \label{eq1}\\
\dot{\boldsymbol{x}}&=&\boldsymbol{f}\left(\boldsymbol{x},V_M\right), \label{eq2}
\end{eqnarray}
where $V_M$ and $I$ are the voltage across and current through the system, respectively, $G\left( \boldsymbol{x}, V_M\right)$ is the memductance (memory conductance), $\boldsymbol{x}$ is a vector of $n$ internal state variables, and $\boldsymbol{f}\left(\boldsymbol{x}, V_M \right)$ is a vector function.

\begin{figure}[b]
\centering{\includegraphics[width=80mm,angle=0]{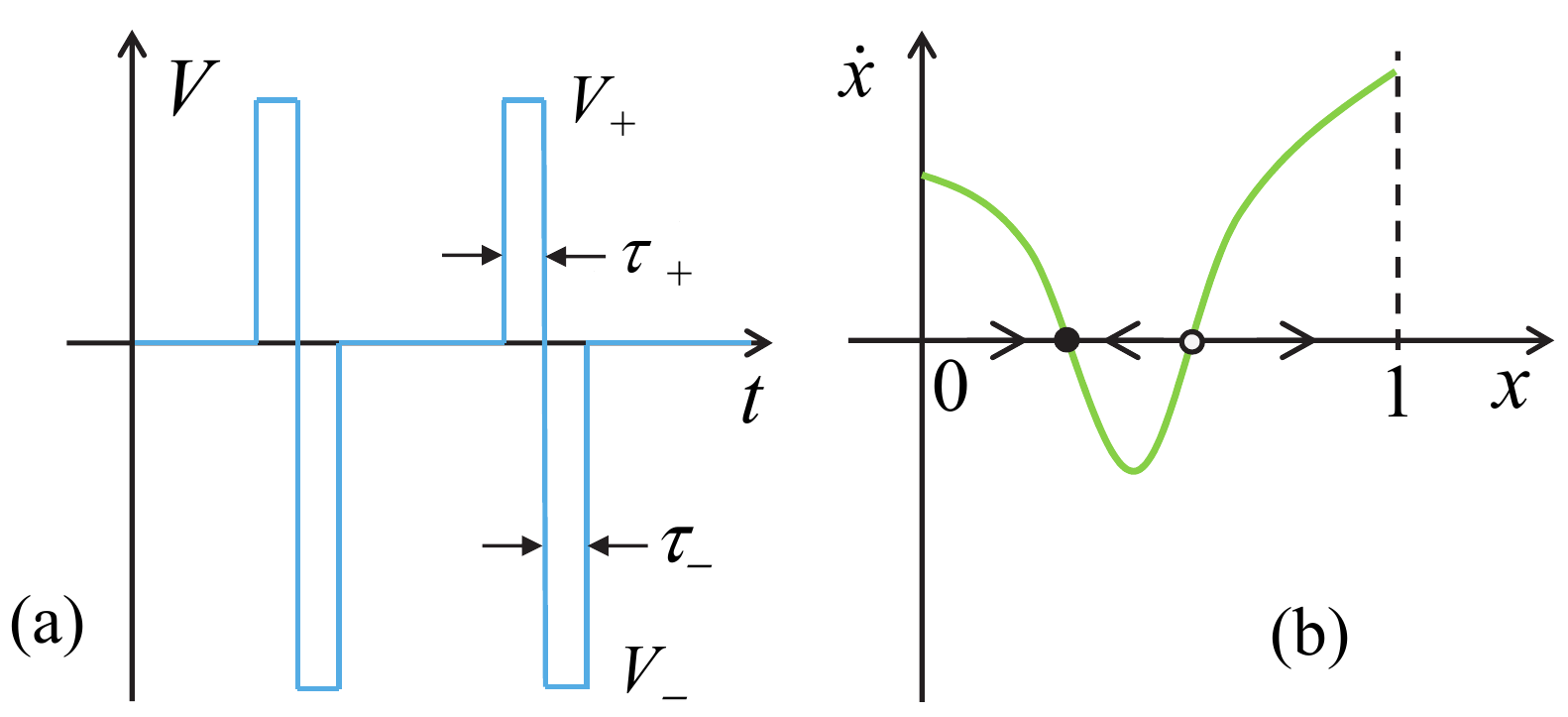}}
\caption{(a) Pulse sequence used in this study. (b) Schematic diagram demonstrating a stable (solid circle) and an unstable (unfilled circle) fixed point.}\label{fig:1}
\end{figure}

In principle, memristors may have quite interesting behavior from the dynamical system point of view~\cite{strogatz2018nonlinear}.
In fact, very recently we have demonstrated the possibility of stable fixed points  in the time-averaged dynamics of memristors driven by narrow pulses of alternating polarities~\cite{pershin18b} (see Fig.~\ref{fig:1}(a) for schematics of the sequence of pulses).
Additionally, we have identified two broad classes of memristor models with and without a stable fixed point in their dynamics~\cite{slipko2018importance}, and found analytical solutions for the transient dynamics of two types of pulse-driven memristors in the presence of a stable fixed point~\cite{slipko19a}.

As almost all previous results~\cite{pershin18b,slipko2018importance,slipko19a} on fixed points of pulse-driven memristors
are limited to simple models, it is of great interest to  understand the possible occurence of stable fixed points in the dynamics of physical memristors.
In this paper, we consider the dynamics of a TaO memristor subjected to narrow pulses of alternating polarities. Our considerations are based on an accurate physics-based model introduced
in \cite{Strachan13a}, and subsequently adopted in several other studies~\cite{Ascoli17a,Ntinas18a}.
The analysis is simplified by the fact that the TaO memristor model~\cite{Strachan13a} involves a single internal state variable.
Therefore, to find the stable fixed points, it is sufficient to analyze the sign of an effective evolution function ($g(x,V_+,V_-)$ in Eq.~(\ref{eq:4})).
One of our main results is that there exists a region of pulse parameters where the number of stable fixed points is two (see Fig.~\ref{fig:3}).
In addition, analytical expressions are derived for the bifurcation curves separating the regions with different numbers of stable fixed points (Section~\ref{sec:an}).

\begin{figure}[t]
\centering{\includegraphics[width=60mm]{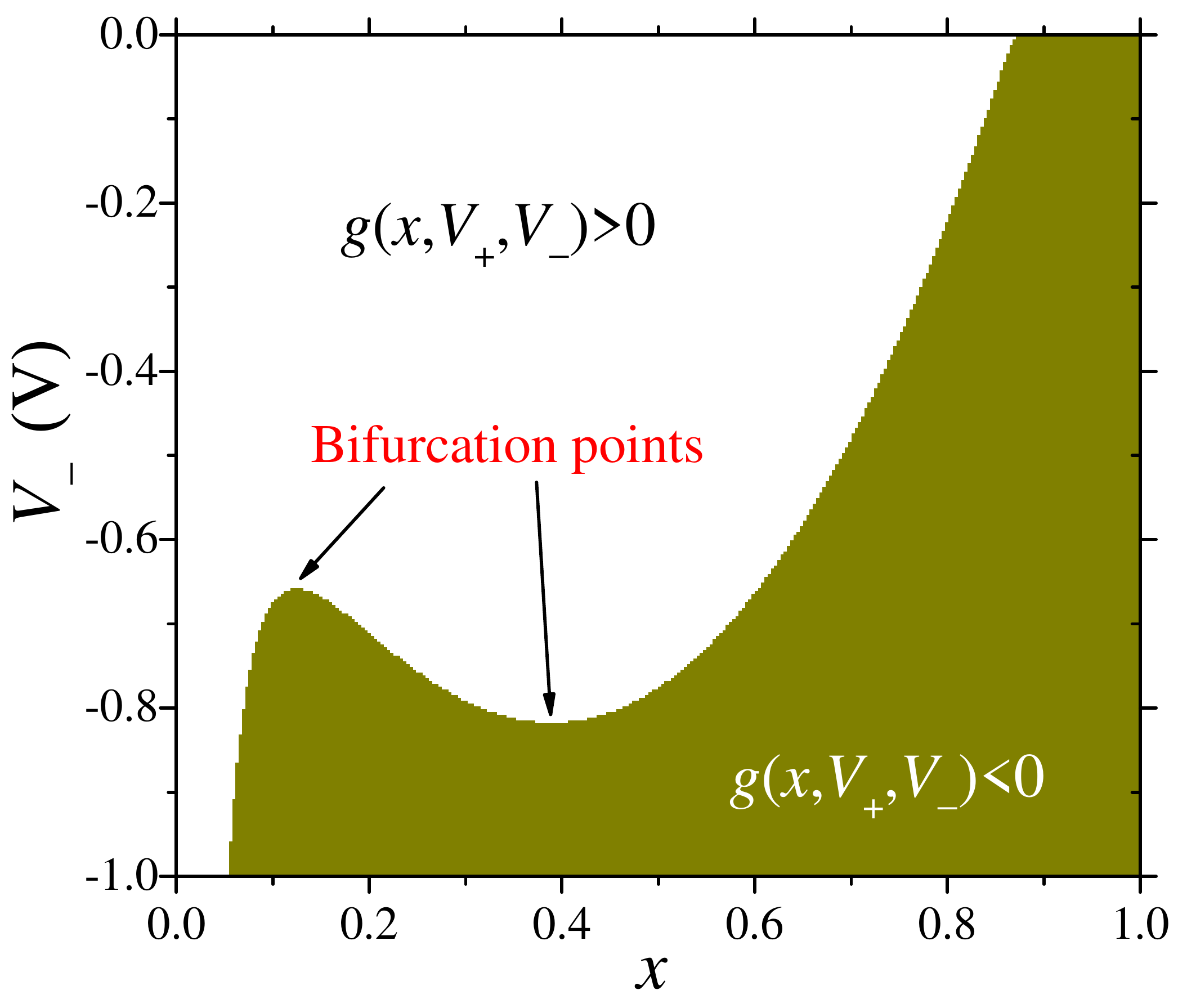}}
\caption{Sign of the evolution function $g(x,V_+,V_-)$  at  $V_+=0.6$~V and $\tau_1=\tau_2$. This function is defined by Eq.~(\ref{eq:4}).}\label{fig:2}
\end{figure}

\section{Model}
According to \cite{Strachan13a}, some TaO memristors are accurately described by equations of the form
\begin{eqnarray}
  I &=& G(x,V_M)V_M \label{eq:1} \;\; , \\
  \dot{x} &=& A \sinh\left(\frac{V_M}{\sigma_{off}}\right)\exp\left( -\frac{x_{off}^2}{x^2}\right)\exp\left( \frac{1}{1+\beta I\;V_M}\right)H(-V_M) \nonumber \\
          &+& B \sinh\left(\frac{V_M}{\sigma_{on}}\right)\exp\left( -\frac{x^2}{x^2_{on}}\right)\exp\left( \frac{I\;V_M}{\sigma_{p}}\right)H(V_M) \;\; , \label{eq:2}
\end{eqnarray}
where the memductance $G(x,V_M)$ is
\begin{equation}\label{eq:3}
  G(x,V_M)=G_Mx +a \exp\left(b\sqrt{|V_M|}\right)(1-x),
\end{equation}
$A$, $B$, $\sigma_{off(on,p)}$, $x_{off(on)}$, $\beta$, $G_M$, $a$, $b$ are constants, and $H(...)$ is the Heaviside step function.
All simulations reported below were performed using the following set of parameter values~\cite{Ascoli17a}:
$A=10^{-10}$~s$^{-1}$, $B=10^{-4}$~s$^{-1}$, $\sigma_{off}=0.013$~V, $\sigma_{on}=0.45$~V, $\sigma_{p}=4\times 10^{-5}$~A$\;$V, $x_{off}=0.4$, $x_{on}=0.06$, $\beta=500$~A$^{-1}$V$^{-1}$, $G_M=0.025$~S, $a=7.2$~$\mu$S, $b=4.7$~V$^{-1/2}$.

Previously, we have shown that stable fixed points (attractors) may  exist in the time-averaged dynamics of driven memristors~\cite{pershin18b,slipko2018importance}. In the case of  first-order memristive systems ($\boldsymbol{x}$ is a scalar in Eqs.~(\ref{eq1})--(\ref{eq2})), the fixed points can be  determined as the time-independent solutions of the following equation~\cite{slipko2018importance}
\begin{equation}
\dot{\bar{x}}(t)=\frac{1}{T}\left(f(\bar{x},V_+)\tau_++f(\bar{x},V_-)\tau_-\right)\equiv g(\bar{x},V_+,V_-). \label{eq:4}
\end{equation}
Here, $\bar{x}(t)=\frac{1}{T}\int\limits_{t}^{t+T} x(\tau)\textnormal{d}\tau$ is the internal state variable averaged over the pulse period $T$, $V_+$ and $V_-$ are the pulse amplitudes, $\tau_+$ and $\tau_-$ are the pulse widths (for their definition, see Fig.~\ref{fig:1}(a)). A fixed point $x_a$ is stable if
\begin{equation}
\left.\frac{\partial g(\bar{x},V_+,V_-)}{\partial \bar{x}}\right|_{\bar{x}=x_a}<0,  \label{eq:5}
\end{equation}
and unstable if
\begin{equation}
\left.\frac{\partial g(\bar{x},V_+,V_-)}{\partial \bar{x}}\right|_{\bar{x}=x_a}>0.  \label{eq:5a}
\end{equation}

In what follows, it is assumed that $\tau_+=\tau_-$, $V_+>0$, and $V_-<0$. Moreover, henceforth, we omit the bar over $x$, considering only the time-averaged internal state variable.

\section{Numerical results} A bifurcation can be defined as a qualitative change in the behavior of a dynamical system with a change of parameters. In the present paper we study how a change in the pulse amplitudes (the parameters $V_+$ and $V_-$) influences the fixed points of driven memristors. At fixed $V_+$ and $V_-$, the equation $\dot{x}=g(x,V_+,V_-)$
represents a vector field on the line, defining the velocity  $\dot{x}$ at each $x$~\cite{strogatz2018nonlinear}.
To understand the qualitative behavior of driven memristors, it is sufficient to examine the change in sign of $g(x,V_+,V_-)$: the transition from positive to negative indicates a stable fixed point, while the opposite transition indicates an unstable one
(see Fig.~\ref{fig:1}(b)).

\begin{figure*}[t]
(a) \centering{\includegraphics[width=60mm]{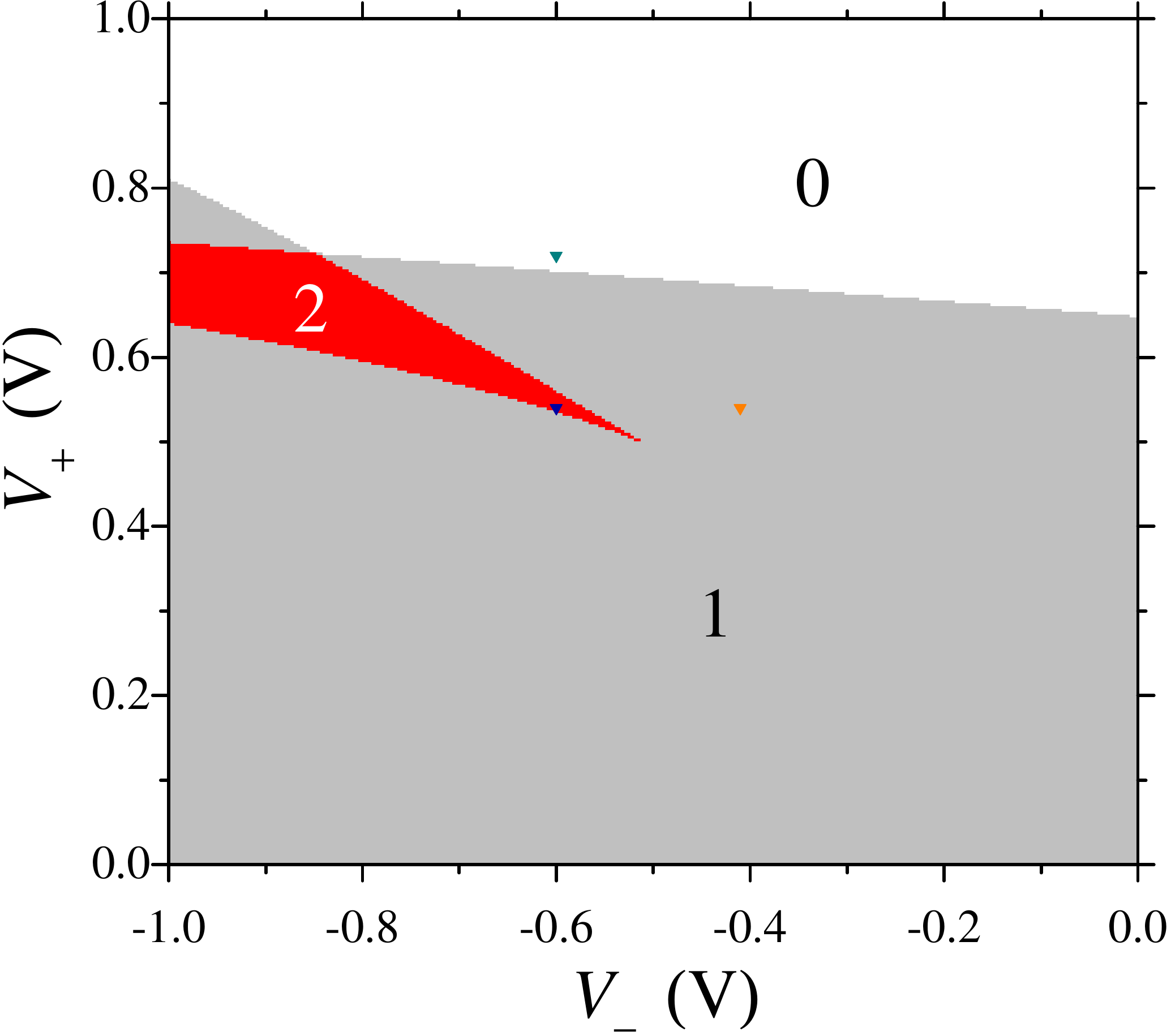}} \;\;\;\;\;\;
(b) \centering{\includegraphics[width=70mm]{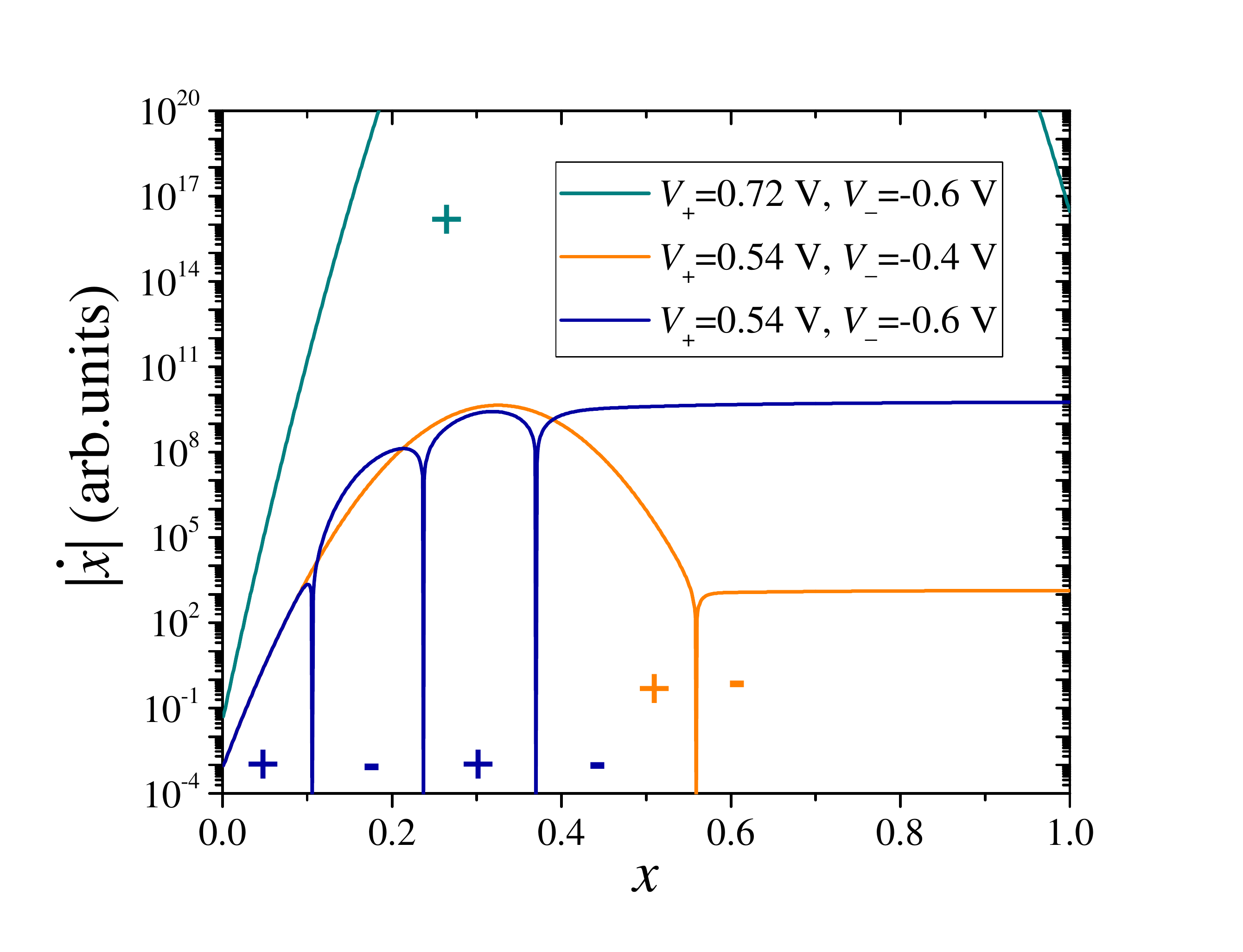}}
\caption{(a) Number of stable fixed points, $N_{st}$, as a function of  pulse amplitudes.  (b) Examples of $g(x,V_+,V_-)$ for selected values of $V_+$ and $V_-$ corresponding to $N_{st}=0,1,2$. Small triangles in (a) correspond to the curves in (b). Pluses and minuses in (b) denote the sign of $g(x,V_+,V_-)$. Here,  $|g(x,V_+,V_-)|$ is shown in a logarithmic scale since $g(x,V_+,V_-)$ spans many orders of magnitude. }\label{fig:3}
\end{figure*}

Fig.~\ref{fig:2} shows the sign of $g(x,V_+,V_-)$ as a function of $x$ and $V_-$ at a fixed value of $V_+=0.6$~V. One can see that
at $V_-< -0.817$~V there exists a single fixed point in the vicinity of $x\approx 0.05$.
As $V_-$ increases, an unstable fixed point and a stable fixed point
are born at $V_-\approx -0.817$~V and shift in opposite directions (a supercritical saddle node bifurcation). At $V_-\approx -0.657$~V, the unstable fixed point born at $V_-\approx -0.817$~V and a stable fixed point located at smaller values of $x$ annihilate each other (a subcritical saddle node bifurcation).

As there are two control parameters involved, $V_+$ and $V_-$, it is worthwhile to analyze the qualitative properties of
the system in a two-dimensional region of these parameters. For this purpose, in Fig.~\ref{fig:3}(a) we plot the number of stable fixed points, $N_{st}$, as a function of
 $V_+$ and $V_-$  (here, each data point was found by a numerical analysis of the zeros of $g(x,V_+,V_-)$).
 Fig.~\ref{fig:3}(a) demonstrates that within the selected domain of pulse amplitudes,  $N_{st}$ ranges between 0 and 2. Fig.~ \ref{fig:3}(b) illustrates the shape of the effective evolution function $g(x,V_+,V_-)$ in various regions of Fig.~\ref{fig:3}(a). For instance, as the leftmost curve $V_+=0.72$~V, $V_-=-0.6$~V is always positive, $N_{st}=0$. $N_{st}=1$ at $V_+=0.54$~V, $V_-=-0.4$~V since the corresponding curve changes its sign once (from $+$ to $-$). The properties of
the $V_+=0.54$~V, $V_-=-0.6$~V curve indicate the existence of two stable and one unstable fixed points.

\begin{figure}[tb]
\centering{\includegraphics[width=70mm]{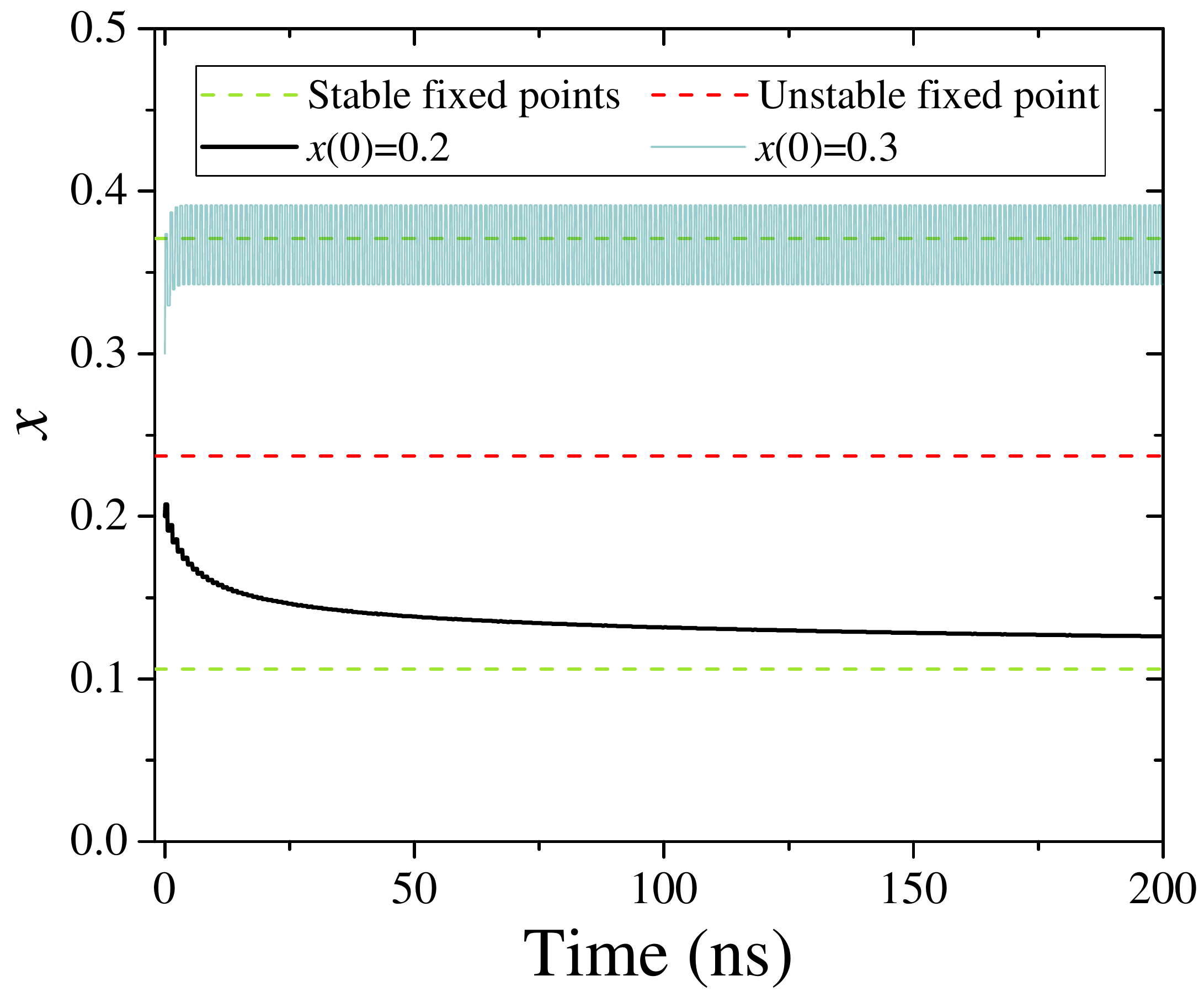}}
\caption{Time dependence of the internal state variable $x$ for two selected initial conditions. This plot demonstrating two distinct basins of attraction was obtained using the following set of parameter values: $V_+=0.54$~V, $V_-=-0.6$~V, $T=1$~ns, $\tau_+=\tau_-=0.1T$ in $x(t=0)=0.2$ calculation, and $\tau_+=\tau_-=0.02T$ in $x(t=0)=0.3$ calculation.}\label{fig:4}
\end{figure}

Finally, we simulate the memristor dynamics for the most interesting case
of $N_{st}=2$ by numerically solving Eqs.~(\ref{eq:1})--(\ref{eq:2}).
From the analysis of the zeros of $g(x,V_+,V_-)$, we have identified the locations of the stable and unstable fixed points, which are indicated by the horizontal dashed lines in Fig.~\ref{fig:4}.
The unstable fixed point at $x\approx 0.237$ separates the basins of attraction of the stable fixed points at $x_1\approx 0.106$ and $x_1\approx 0.371$.
Our numerical modeling shows that when the initial condition (such as $x(0)=0.2$ in Fig.~\ref{fig:4}) belongs to the attraction basin of $x_1$, the system trajectory asymptotically approaches $x_1$.
Similarly, when $x(0)$ belongs to the attraction basin of the second attractor, the trajectory approaches $x_2$.
The oscillations
 about $x_2$ are related to the strong nonlinearity of the effective evolution function (see
the $V_+=0.54$~V, $V_-=-0.6$~V curve $3$ in Fig.~\ref{fig:3}(b)), and can be reduced by using shorter pulses.

\section{Analytical analysis} \label{sec:an}

\begin{figure}[b]
\centering{\includegraphics[width=70mm]{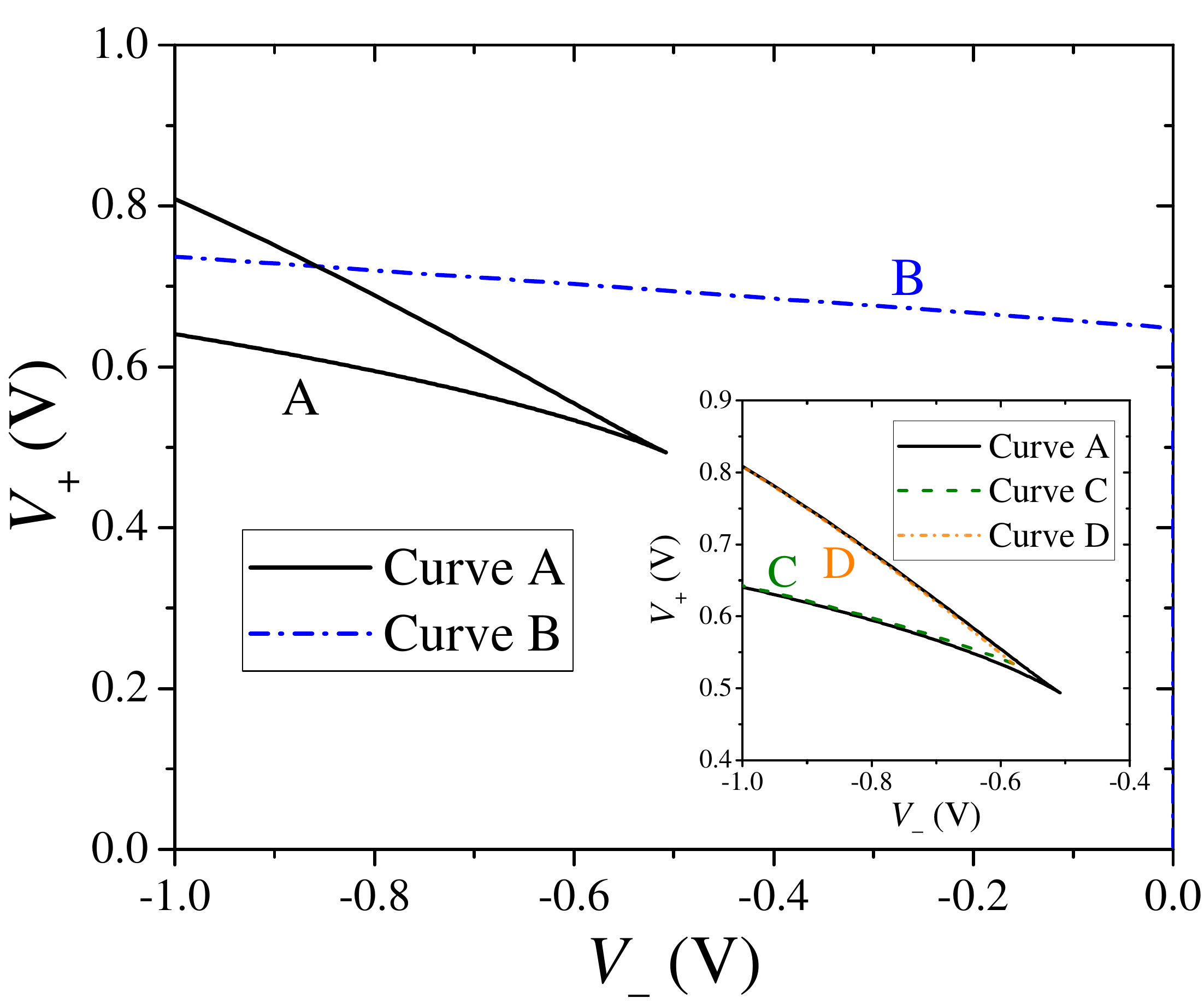}}
\caption{Analytically found bifurcation curves A and B. Inset: curves C and D represent nonparametric expressions for two branches of A.}\label{fig:5}
\end{figure}

In this section, we develop an analytical understanding of Fig.~\ref{fig:3}(a). Our specific goal is to derive approximate analytical expressions for the bifurcation curves.
To simplify the presentation, Fig.~\ref{fig:5} exhibits the curves derived below, which are denoted by A, B, C, and D. For most of their length, curve A separates the
regions $N_{st}=2$ and $N_{st}=1$, while curve $B$ separates the regions $N_{st}=1$ and $N_{st}=0$ (compare Fig.~\ref{fig:5} to Fig.~\ref{fig:3}(a)).
Curves C and D (shown in the inset) represent nonparametric expressions for two branches of A.

Before starting the derivation, recall that at a stable fixed point, the right-hand side of Eq.~(\ref{eq:4}) equals zero. This leads to the equation
\begin{equation}\label{eq:an_1}
  \frac{V_+^2}{\sigma_p}G(x,V_+)+\frac{x_{off}^2}{x^2}=\gamma+\frac{x^2}{x^2_{on}}+\frac{1}{1+\beta G(x,V_-)V_{-}^2},
\end{equation}
where
\begin{equation}\label{eq:an_2}
  \gamma=\textnormal{ln}\frac{ A \tau_-\sinh\left(\frac{|V_-|}{\sigma_{off}}\right)}{B \tau_+\sinh\left(\frac{V_+}{\sigma_{on}}\right)}.
\end{equation}
As Eq.~(\ref{eq:an_1}) is a fifth-order polynomial equation with respect to $x$, its roots can be found only numerically.
We analyzed the contributions of the different terms in Eq. (\ref{eq:an_1}) and found that the last term in Eq.~(\ref{eq:an_1}) is typically small.
Since no analytical solution of Eq.~(\ref{eq:an_1}) exists, in what follows the last term in  Eq.~(\ref{eq:an_1}) is neglected.
The condition of equal derivatives of both sides of Eq.~(\ref{eq:an_1}) provides the second equation for the creation or annihilation of a fixed points pair (in Section~\ref{sec:curve C} below, the geometric meaning of this condition will be discussed in detail).
Thus, our analysis is based on the following set of equations:
\begin{eqnarray}
  \frac{V_+^2}{\sigma_p}G(x,V_+)+\frac{x_{off}^2}{x^2}&=&\gamma+\frac{x^2}{x^2_{on}},
  \label{eq:an_3} \\
  \frac{V_+^2}{\sigma_p}G'_x(x,V_+)-\frac{2x_{off}^2}{x^3}&=&\frac{2x}{x^2_{on}}.
  \label{eq:an_5}
\end{eqnarray}

\subsection{Curve A}\label{sec:curve A}

For the sake of convenience, we rewrite Eqs.~(\ref{eq:an_3}) and (\ref{eq:an_5}) as
\begin{eqnarray}
  \Gamma x+\frac{x_{off}^2}{x^2}=\tilde{\gamma}+\frac{x^2}{x^2_{on}}\label{eq:anN_1}, \\
  \Gamma=\frac{2x_{off}^2}{x^3}+\frac{2x}{x^2_{on}},\label{eq:anN_2}
\end{eqnarray}
where  two  auxiliary variables $\Gamma$ and $ \tilde{\gamma}$ have been introduced:
\begin{eqnarray}
  \Gamma&=&\frac{V_+^2}{\sigma_p}\left(G_M-a e^{b\sqrt{V_+}} \right),\label{eq:anN_3} \\
  \tilde{\gamma}&=&\gamma - \frac{V_+^2}{\sigma_p} a e^{b\sqrt{V_+}}. \label{eq:anN_4}
\end{eqnarray}
Note that $\Gamma=\Gamma(V_+)$ and $\tilde{\gamma}=\tilde{\gamma}(V_+,V_-)$. In principle, $x$ can be eliminated from the
system (\ref{eq:anN_1}) and (\ref{eq:anN_2}); however, the resulting relation between $V_+$ and $V_-$ is cumbersome and inconvenient
for further analysis. It is much more convenient to develop a parametric representation for the bifurcation curve A, where $x$ is considered as a parameter.
Substituting Eq.~(\ref{eq:anN_2}) into  Eq.~(\ref{eq:anN_1}), we can simplify  Eq.~(\ref{eq:anN_1}) to
\begin{equation}\label{eq:anN_5}
 \tilde{\gamma}=\frac{3x_{off}^2}{x^2}+\frac{x^2}{x^2_{on}}
\end{equation}
that, together with Eq.~(\ref{eq:anN_2}), can be considered as a parametric form of the $V_{+}=V_{+}(V_{-})$ curve.

An interesting common feature of Eqs.~(\ref{eq:anN_2}) and (\ref{eq:anN_5}) is a singular point $x_c$ where
\begin{equation}\label{eq:anN_6}
  \Gamma ' (x_c)=\tilde{\gamma}' (x_c)=0.
\end{equation}
A simple calculation shows that this point is located at $x_c=\sqrt[4]{3}\sqrt{x_{on}x_{off}}\approx 0.204$.  One can show that this singular point is a turning point (a cusp), namely, a point on a curve where a moving point on the curve must start to move backward.
At the cusp, two parts of a curve meet tangentially to each other~\cite{pogorelov1959}.

To find $V_+(x)$, we note that in a zero-order approximation one can neglect the exponential term in (\ref{eq:anN_3}) to get
\begin{equation}\label{eq:anN_7}
  V_+^{(0)}=\sqrt{\frac{\Gamma(x)\sigma_p}{G_M}}.
\end{equation}
Rewriting (\ref{eq:anN_3}) as
\begin{equation}\label{eq:anN_9}
  V_+=\sqrt{\frac{\Gamma(x)\sigma_p}{G_M}}\frac{1}{\sqrt{1-\frac{a}{G_M}e^{b\sqrt{V_+}}}}
\end{equation}
and using Eq. (\ref{eq:anN_7}), we finally obtain
\begin{equation}\label{eq:anN_10}
  V_+=\sqrt{\frac{2\sigma_p}{G_M}\left( \frac{x_{off}^2}{x^3}+\frac{x}{x^2_{on}} \right)}\left(1+\frac{a}{2G_M}e^{b\sqrt[4]{\frac{2\sigma_p}{G_M}\left( \frac{x_{off}^2}{x^3}+\frac{x}{x^2_{on}} \right)}} \right).
\end{equation}
With the help of Eqs. (\ref{eq:an_2}) and (\ref{eq:an_3}), we arrive at the equation
\begin{equation}\label{eq:anN_11}
 V_-=-\sigma_{off}\textnormal{arcsinh} \left[\frac{B\tau_+}{A\tau_-}\sinh\left( \frac{V_+(x)}{\sigma_{on}}\right)\exp
   \left\{\frac{x_{off}^2}{x^2}-\frac{x^2}{x^2_{on}}+\frac{V_+^2(x)}{\sigma_p} G(x,V_+(x))  \right\}.
  \right]
\end{equation}

Eqs. (\ref{eq:anN_10}) and (\ref{eq:anN_11}) define curve A parametrically (as a function of $x$), and were used to obtain curve A in Fig.~\ref{fig:5}. In particular, by substituting the known value of  $x_c$  into the right-hand sides of Eqs.~(\ref{eq:anN_10}) and (\ref{eq:anN_11}), we find the critical values of the voltage pulse amplitudes $V_+^{(c)}\approx 0.49$~V and $V_-^{(c)}\approx -0.51$~V.
Note that these values correspond to the cusp in Fig.~\ref{fig:3}(a) or Fig.~\ref{fig:5}.

\subsection{Curve B}\label{sec:curve B}
Next, we derive curve B. This curve corresponds
to the transition of a stable fixed point through the right boundary $x=1$ (see Fig.~\ref{fig:6}(e)). Substituting $x=1$ into Eq. (\ref{eq:an_3}) and taking into account that $G_M(1,V_+)=G_M$ we find
\begin{equation}\label{eq:an_13}
  \gamma=\frac{V_+^2}{\sigma_p}G_M-\frac{1}{x_{on}^2},
\end{equation}
where a small term $x_{off}^2=0.16$ was omitted as it is of the order of the previously omitted last term in Eq.~(\ref{eq:an_1}). Taking into account the definition of $\gamma$ (see Eq.~(\ref{eq:an_2})), we obtain
\begin{equation}\label{eq:an_14}
V_-=-\sigma_{off}\textnormal{arcsinh}\left[\frac{B\tau_+}{A\tau_-}\sinh\left( \frac{V_+}{\sigma_{on}}\right)\exp\left[ \frac{V_+^2}{\sigma_p}G_M -\frac{1}{x_{on}^2}\right] \right].
\end{equation}
This equation describes curve B, and is presented in Fig.~\ref{fig:5}.

\subsection{Curve C}\label{sec:curve C}
\begin{figure*}[t]
\centering{(a)\includegraphics[width=45mm]{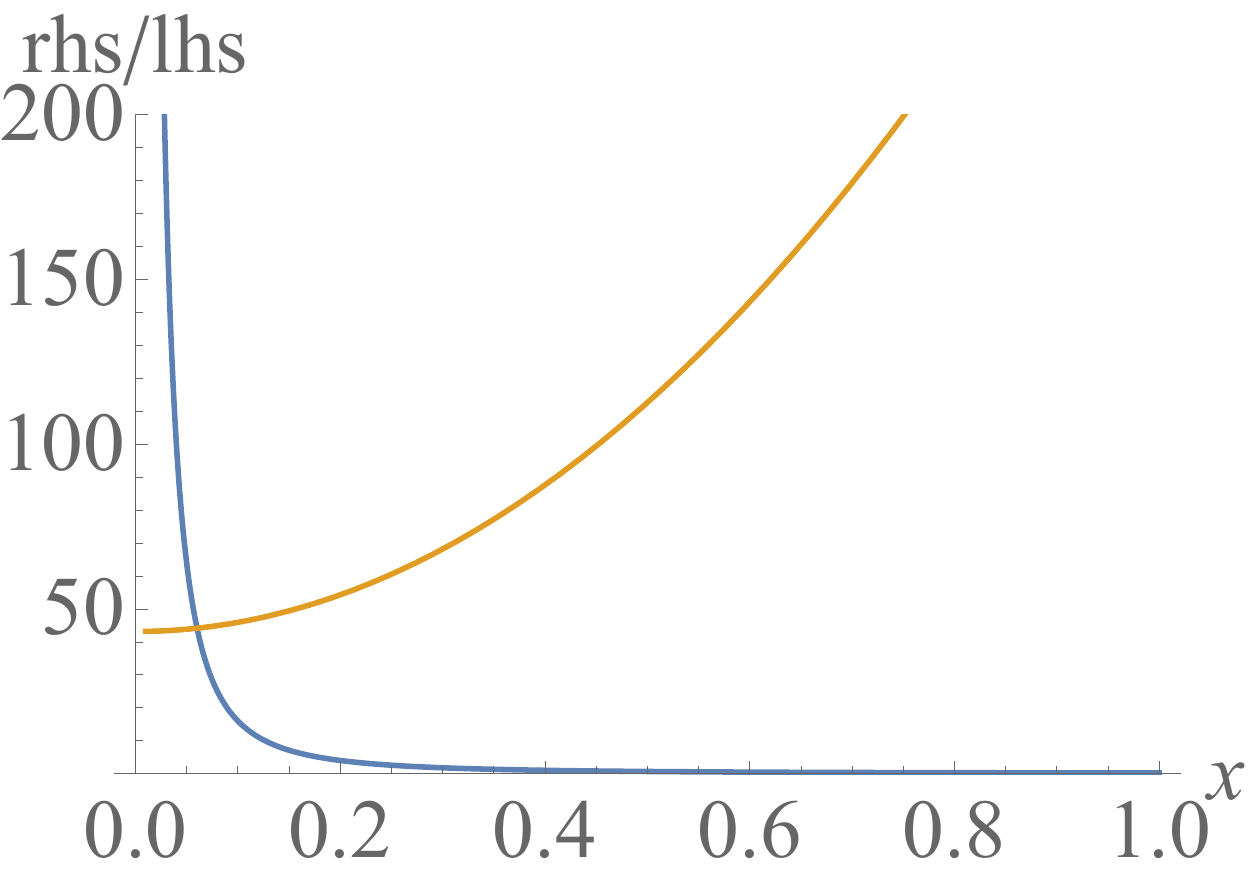}\;\;(b)\includegraphics[width=45mm]{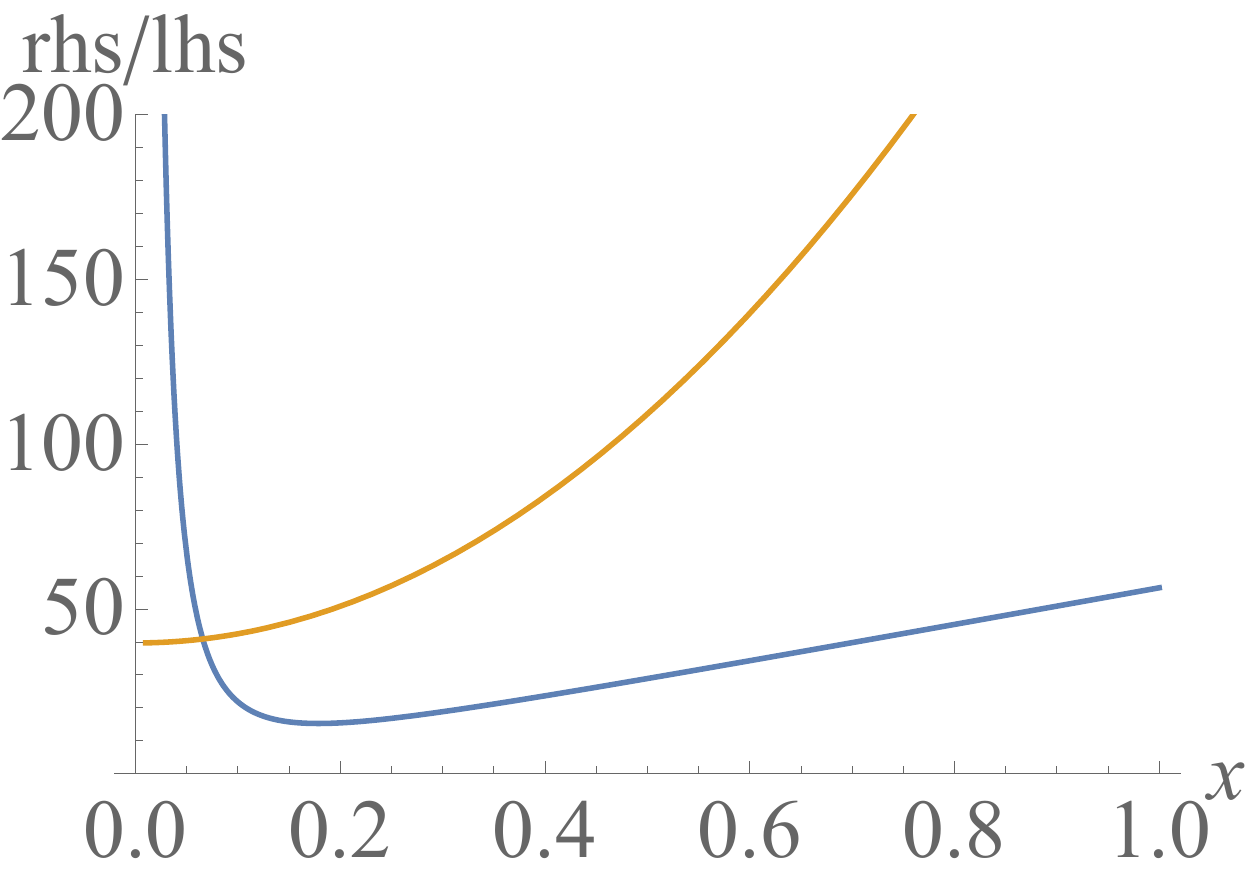}\;\;(c)\includegraphics[width=45mm]{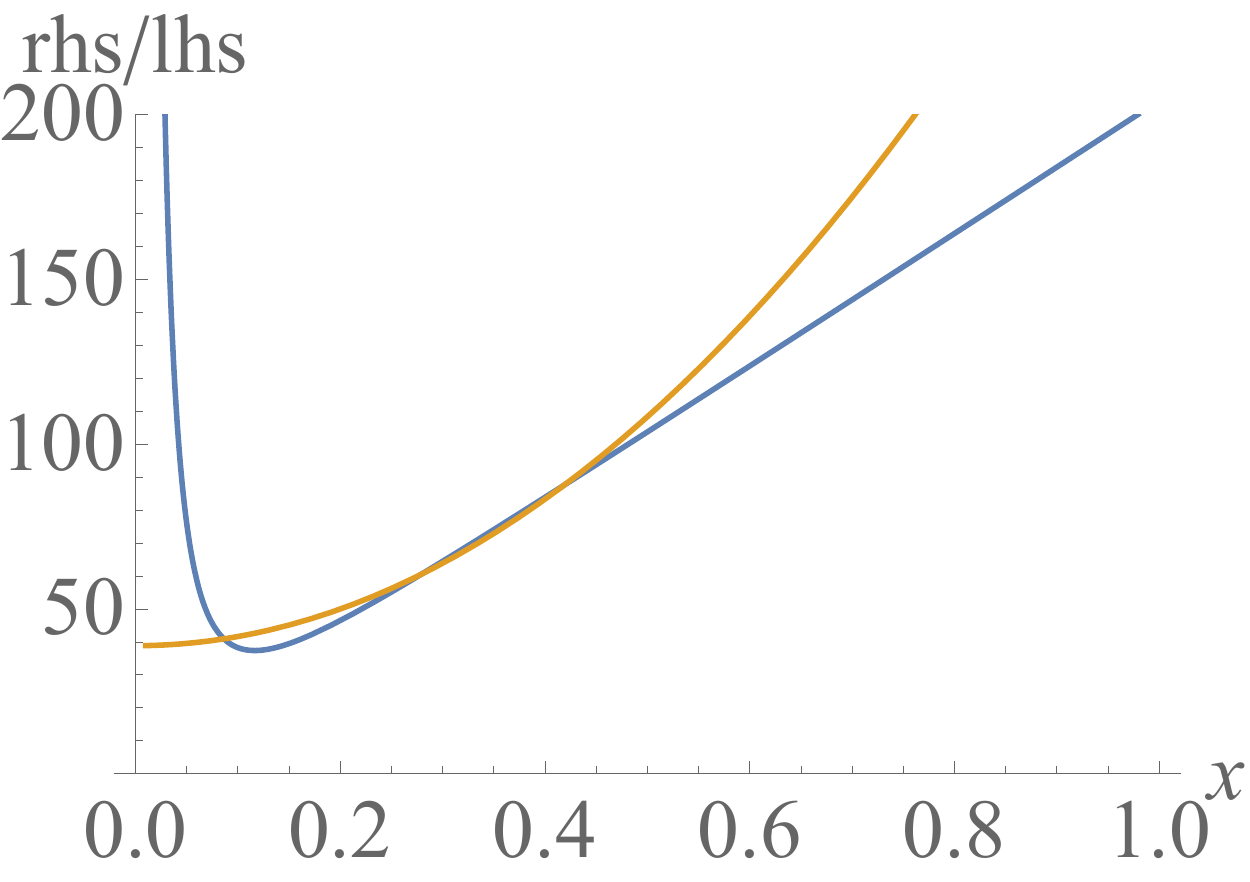}}
\centering{(d)\includegraphics[width=45mm]{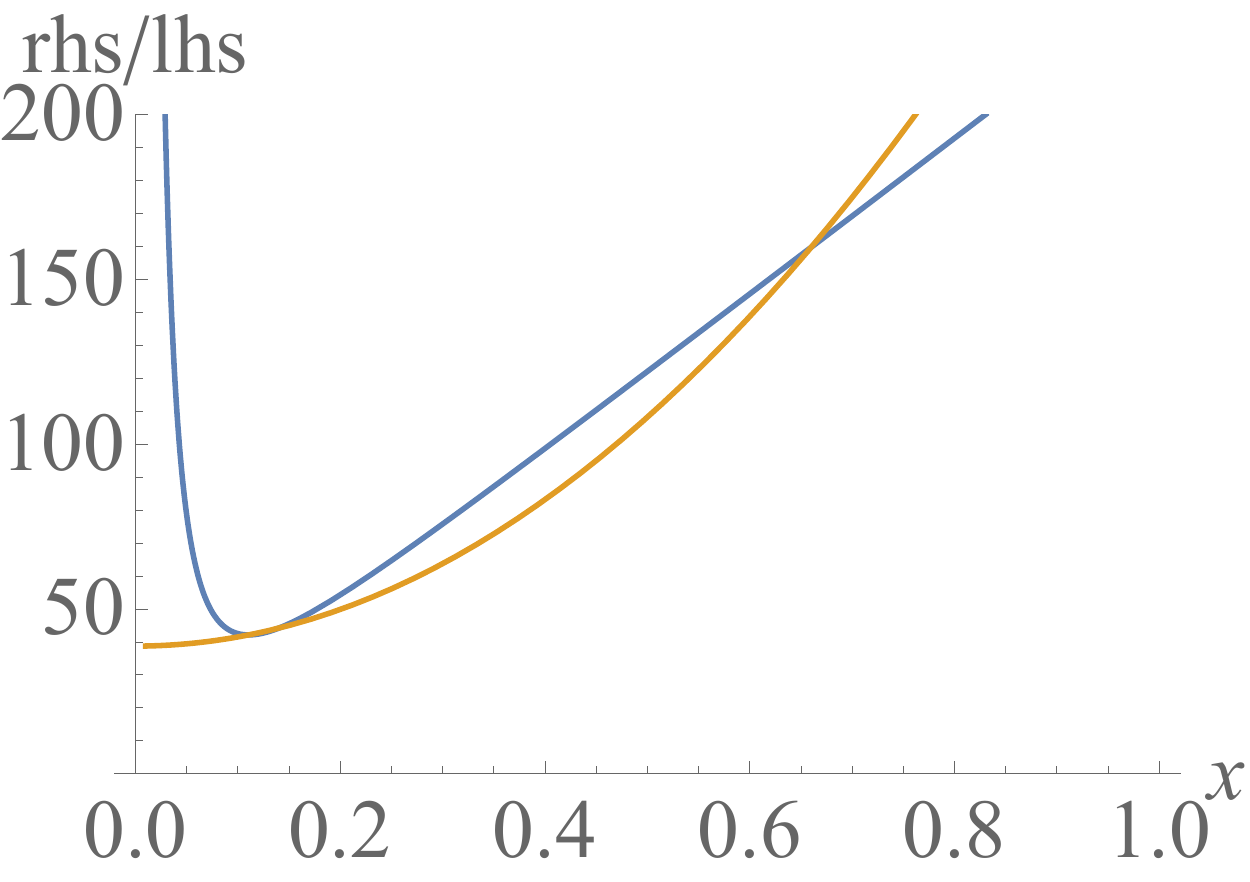}\;\;(e)\includegraphics[width=45mm]{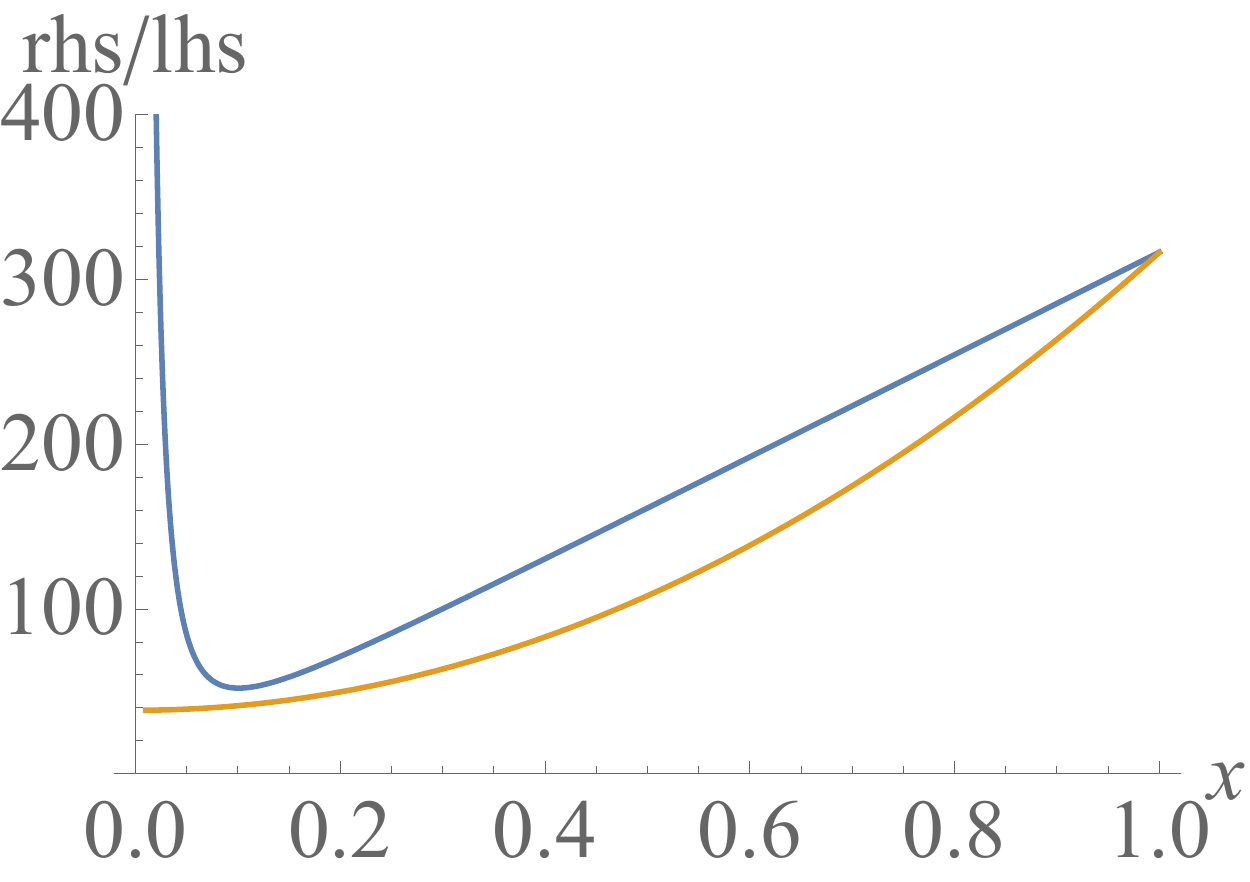}}
\caption{Left-hand side (blue line) and right-hand side (orange line) of Eq.~(\ref{eq:an_3}) at (a) $V_+=0.01$~V, (b) $V_+=0.3$~V, (c) $V_+=0.5714$~V, (d) $V_+=0.6197$~V, and (e) $V_+=0.71136$~V. All graphs were obtained at $V_-=-0.7$~V.}\label{fig:6}
\end{figure*}

At small values of $V_+<0.023$~V, the left-hand side of Eq.~(\ref{eq:an_3}) is a decreasing function of $x$, while its right-hand side is always increasing (for $x\in [0,1]$).
In this case one can show that Eq.~(\ref{eq:an_3}) has a single root practically for all values of $V_-$ in the interval $0\geq V_-\geq -1$~V.
This situation is presented graphically in Fig.~\ref{fig:6}(a), where the root corresponds to the point of intersection of the  curves.
As $V_+$ increases above $0.023$~V, the number of roots stays initially the same (one), although the curve representing the left-hand side of Eq.~(\ref{eq:an_3}) becomes non-monotonic (Fig.~\ref{fig:6}(b)).
At a certain  point, $V_+\approx 0.5$~V and $V_-\approx -0.5$~V, the curves become tangent to one another (Fig.~\ref{fig:6}(c)).
This boundary situation between having only one root and having three roots of Eq.~(\ref{eq:an_3}) provides the functional relation between $V_+$ and $V_-$ at the boundary (curve A).

To derive C, which is the lower branch of A, we make the following simplification. Assuming that $x_{off}^2/x^2$ is small at the point where C and D are tangent to one another, this term is neglected in Eq.~(\ref{eq:an_3}). This leads to the quadratic equation
\begin{equation}\label{eq:an_6}
\frac{V_+^2}{\sigma_p}G(x,V_+)=\gamma+\frac{x^2}{x^2_{on}}
\end{equation}
or
\begin{equation}\label{eq:an_7}
\frac{x^2}{x^2_{on}} -\frac{V_+^2}{\sigma_p}(G_M-ae^{b\sqrt{V_+}})x+\left(\gamma-\frac{V_+^2a}{\sigma_p}e^{b\sqrt{V_+}} \right)  =0.
\end{equation}
The condition of zero discriminant (corresponding to the point of interest) results in the equation
\begin{equation}\label{eq:an_8}
  \gamma=\left[ \frac{V_+^2 x_{on}}{2\sigma_p}(G_M-ae^{b\sqrt{V_+}}) \right]^2+\frac{V_+^2}{\sigma_p}ae^{b\sqrt{V_+}}.
\end{equation}
Using Eqs.~(\ref{eq:an_8}) and (\ref{eq:an_2}) one finds $V_-\left( V_+\right)$:
\begin{equation}\label{eq:an_9}
 V_-=-\sigma_{off}\textnormal{arcsinh} \left[\frac{B\tau_+}{A\tau_-}\sinh\left( \frac{V_+}{\sigma_{on}}\right)\exp
  \left\{ \frac{V_+^4x_{on}^2}{4\sigma_p^2}\left(G_M-ae^{b\sqrt{V_+}}\right)^2+\frac{V_+^2}{\sigma_p}ae^{b\sqrt{V_+}} \right\}
  \right].
\end{equation}
This equation describes curve C, and is presented in the inset in Fig.~\ref{fig:5}.

\subsection{Curve D}
 When $V_+>0.81$~V and $0\geq V_-\geq -1$~V, Eq.~(\ref{eq:an_3}) has no roots in the interval $x\in [0,1]$. As $V_+$ decreases, two roots may appear close to the minimum of $(V_+^2/ \sigma_p) G(x,V_+)+x_{off}^2/x^2$ (see Eq.~(\ref{eq:an_3})), and one root may appear closer to 1. It is important to note that in the vicinity of the minimum, the term $x_{off}^2/x^2$ can not be neglected.

One can see that the upper boundary of the region $N_{st}=2$ corresponds to the condition when the two curves representing Eq.~ (\ref{eq:an_3}) touch at a single point (see Fig.~\ref{fig:6}(d)). The position of the minimum can be found from
\begin{equation}\label{eq:an_10}
  \frac{V_+^2}{\sigma_p}G'_x(x,V_+)-\frac{2x_{off}^2}{x^3}=0,
\end{equation}
which gives
\begin{equation}\label{eq:an_11}
  x_{min}=\sqrt[3]{\frac{2x_{off}^2\sigma_p}{V_+^2\left( G_M-ae^{b\sqrt{V_+}}\right)}}.
\end{equation}
Substituting Eq.~(\ref{eq:an_11}) into Eq.~(\ref{eq:an_3}) we find curve D:
\begin{equation}\label{eq:an_12}
V_-=-\sigma_{off}\textnormal{arcsinh}\left[\frac{B\tau_+}{A\tau_-}\sinh\left( \frac{V_+}{\sigma_{on}}\right)\exp\left[ \gamma \right] \right],
\end{equation}
where $\gamma$ is defined by Eq.~(\ref{eq:an_3}) with $x=x_{min}$ from Eq.~(\ref{eq:an_11}).
Curve D is described by Eq.~(\ref{eq:an_12}) and is presented in Fig.~\ref{fig:5} inset.

\section{Conclusion} In this paper, we have discovered that two stable fixed points may exist simultaneously in the time-averaged dynamics of a pulse-driven memristor.
We have identified and completely determined the parameters of the sequence of pulses that lead to a specific number of stable fixed points, a number which is either 0, 1, or 2.
As the parameters of the sequence of pulses can be easily adapted/controlled in a circuit or experiment, our findings open up a novel approach to the use of memristor in a range of applications, from computing to emerging neuromorphic circuits.

Our very interesting {\it theoretical} results are based, of course, on the specific memristor model~\cite{Strachan13a} in which  the memristor evolution function $f(x,V_M)$  is a non-separable and non-monotonic function of $x$.
We note that due to these properties,  the theorems formulated in~\cite{slipko2018importance} are not applicable in the present case.
Experimental verification of our results is both possible and desirable.

\section*{References}

\bibliographystyle{IEEEtran}
\bibliography{IEEEabrv,bifurc}

\end{document}